\documentclass[amsmath, amsfonts, prb, showpacs, twocolumn]{revtex4}
\usepackage{graphicx}
\usepackage{epsfig}
\usepackage{bm}
\usepackage{euscript}
\usepackage{dcolumn}
\usepackage{color}
\usepackage{amsmath}
\usepackage{amssymb}
\usepackage[varg]{txfonts}

\def\Tr{\mathrm{Tr}}

\begin{document}

\title{Josephson current noise above $T_c$ in superconducting tunnel junctions}

\author{Alex Levchenko}

\affiliation{School of Physics and Astronomy, University of
Minnesota, Minneapolis, MN, 55455, USA}

\begin{abstract}
Tunnel junction between two superconductors is considered in the
vicinity of the critical temperature. Superconductive fluctuations
above $T_c$ give rise to the noise of the ac Josephson current
although the current itself is zero in average. As a result of
fluctuations, current noise spectrum is peaked at the Josephson
frequency, which may be considered as precursor of superconductivity
in the normal state. Temperature dependence and shape of the
Josephson current noise resonance line is calculated for various
junction configurations.
\end{abstract}

\date{September 12, 2008}

\pacs{74.40.+k, 74.50.+r}

\maketitle

\section{Introduction}

In the vicinity of the critical temperature $T_c$ transport
properties of metals are strongly affected by superconductive
fluctuations. For example, in the temperature region $T-T_c\ll T_c$,
where fluctuations are the most pronounced,  Drude conductivity
acquires noticeable Aslamazov-Larkin, Maki-Thompson, and
density-of-states (DOS) corrections. Many other kinetic and
thermodynamic coefficients such as magnetic susceptibility, heat
conductivity, Hall coefficient, and ultrasonic attenuation are also
modified by fluctuations. One may consult recent book
(Ref.~\onlinecite{Larkin-Varlamov}) for exhaustive overview of
results and literature in this field.

Mostly immediately after the pioneering works on superconductive
fluctuations,~\cite{Aslamazov-Larkin,Maki} it was noticed that
analog of the ac Josephson effect may survive in the normal state
above the critical temperature.~\cite{Kulik,Scalapino} The latter is
also attributed to the formation of fluctuating Cooper pairs.
Indeed, consider weak transparency tunnel junction between two
superconductors. In this case Josephson current is given by
$I_J(t)=I_c\sin(\omega_J t)$, where $\omega_J=2eV$ is the Josephson
frequency and current amplitude $I_c$ is proportional to the product
of superconductive order parameters $\Delta_{L(R)}$, taken from the
left ($L$) and right ($R$) to the contact area. Above the critical
temperature Josephson current vanishes $\langle
I_J\rangle\sim\langle\Delta_L\Delta_R\rangle=0$ since order
parameter is zero in average $\langle\Delta_{L(R)}\rangle\equiv0$.
However, current squared $\langle
I^{2}_{J}\rangle\sim\langle\Delta_{L}\Delta_{R}\Delta_{L}\Delta_{R}\rangle$,
which gives noise of the Josephson current, is apparently not zero
due to nontrivial average $\langle\Delta_{L(R)}\Delta_{L(R)}\rangle$
of space and time fluctuating order parameters. As a consequence,
noise power spectrum $S_J(\omega)$, defined as the Fourier transform
of Josephson current-current correlation function, shows distinctive
peak at the Josephson frequency $\omega_J$, which is experimentally
an accessible effect. The peak height
$S_{\mathrm{max}}=S_{J}(\omega=\omega_{J})$ is a strong function of
$T-T_{c}$, usually some power law, which makes it possible to detect
noise signal in the immediate vicinity of the critical temperature
$T-T_{c}\ll T_{c}$. Although this observation was there for a long
time, the interest to it was recently revived. It was
stressed~\cite{Varlamov-Review,Martin-Balatsky,Dai} that
measurements of the Josephson current noise may be especially
fruitful in studies of the high--temperature superconductivity.
Indeed, whether superconductive pairing fluctuations exist in the
pseudogap regime of the high--$T_c$ materials may be probed by the
Josephson tunneling. Thus, existence of the Josephson effect above
$T_c$ may be thought as the precursor of superconductivity.

So far fluctuations of the Josephson current above the critical
temperature were studied either for the narrow
contacts~\cite{Kulik,Martin-Balatsky,Dai}, taking into account only
temporal fluctuations of the order parameter, or for the mesoscopic
rings~\cite{Dorin,Shimshoni}. We find, however, that in the planar
geometry of the tunnel junction, where spatial variations of the
superconductive order parameter have to be accounted for, peak in
the current noise spectrum is more pronounced, especially, for the
non--symmetric junction configurations. Motivated by the ongoing
experiments~\cite{Reznikov} and possible applications in probing
pseudogap regime of high--$T_c$ materials, we revisit problem of the
Josephson current noise above $T_c$ and study noise in the planar
geometry of a tunnel junction. Within this work we focus on the
temperature range $Gi\ll(T-T_{c})/T_{c}\lesssim1$, where $Gi$ is the
Ginzburg number. In this regime fluctuations can be considered as
small and can be treated in perturbation theory. The natural
expansion parameter, which measures strength of the superconductive
fluctuations, is $Gi\lesssim 1$.

The main results of the present work may be summarized as follows:
(i) For symmetric wide junctions, when both electrodes are in the
fluctuating regime, and contact area $\mathcal{A}$ is large as
compared to the square of the superconductive coherence length,
$\mathcal{A}\gtrsim\xi^{2}_{o}$, Josephson current noise spectrum
$S_{J}(\omega)$ has a Lorentzian--like shape. The peak height scales
in temperature as
$S_{\mathrm{max}}\propto\left(\frac{T_{c}}{T-T_{c}}\right)^{2}$ and
depends quadratically on both tunnel conductance of the junction
$g_{T}$ and the Ginzburg number $Gi$. For the lowest temperature
$T-T_{c}=Gi\, T_{c}$, which is allowed by the applicability of the
perturbation theory, strength of the noise is given by
$S_{\mathrm{max}}=(\pi/64)\big(g^{2}_{T}T_{c}/e^{2}\big)
\big(\xi^{2}_{o}/\mathcal{A}\big)$. Of course, experimentally, noise
is maximal right at the transition $T=T_c$; however, in this case it
is very difficult to make any quantitative predictions
theoretically. Thus, $S_{\mathrm{max}}$ gives an order of magnitude
estimate. (ii) For the narrow, $\xi^{2}_{o}\gtrsim\mathcal{A}$,
symmetric junctions we find also a Lorentzian--like shape of
$S_J(\omega)$, which is again quadratic in both $g_{T}$ and $Gi$;
however, temperature dependence of the peak height is different
$S_{\mathrm{max}}\propto\frac{T_{c}}{T-T_{c}}$. The estimate for the
noise power at the most vicinity of the transition is
$S_{\mathrm{max}}=(Gi/8\pi)(g^{2}_{T}T_{c}/e^{2})$. (iii) In the
case of non--symmetric junctions, when one electrode is already
superconducting while another is fluctuating, noise has Lorentzian
form. The temperature dependence for the peak height in this case is
the same as for wide symmetric junction, which, however, appears
already in the first order of the Ginzburg number and contains large
prefactor $\ln^{2}(\Delta_{S}/T_{c})$ (where $\Delta_{S}$ is the
superconductive gap). (iv) Corrections to the current noise above
$T_{c}$ are not exhausted by the Josephson current contribution
only. In addition, superconductive fluctuations deplete
normal--metal DOS at the Fermi energy, which changes tunnel
conductance. The latter translates into the current noise correction
$S_{\mathrm{DOS}}(\omega)$ via fluctuation dissipation theorem
(FDT). This effect is linear in $g_T$ and $Gi$, logarithmic in
temperature
$S_{\mathrm{DOS}}\propto\ln\left(\frac{T_{c}}{T-T_{c}}\right)$, and
has an opposite sign as compared to the Josephson current
contribution.

The rest is organized as follows: in the next section
(Sec.~\ref{sec-Formalism}) we present in a concise form our
technical method, Keldysh nonlinear $\sigma$-model, which will be
used through out the paper in calculation of the current noise
power. This formalism was elaborated in Refs.~\onlinecite{KA} and
\onlinecite{FLS}, and found to be very useful and powerful in many
applications. In the Sec.~\ref{sec-Noise} we calculate
density-of-states and Josephson current contributions to the noise
spectrum above $T_{c}$. The results of the work together with
further discussions are summarized in the Sec.~\ref{sec-Summary}.
Number of technical points are delegated to the
Appendixes~\ref{sec-Fluctuations}--\ref{sec-Integrals}.

\section{Formalism}\label{sec-Formalism}

Consider voltage biased tunnel junction of two superconductors above
the critical temperature. Within $\sigma$--model formalism tunneling
between $L$ and $R$ reservoirs of a junction is described by the
action,
\begin{equation}\label{S-T}
i\mathbb{S}_{T}[V]=\frac{\pi
g_{T}}{4e^{2}}\Tr\left\{e^{i\check{\Xi}\check{V}}\check{Q}_{L}
e^{-i\check{\Xi}\check{V}}\check{Q}_{R}\right\},
\end{equation}
where $g_{T}$ is the junction tunnel conductance and
$\check{Q}_{L(R)}$ are the Green's functions describing electron
system in the electrodes (hereafter $\hbar=k_{B}=1$).  Both
$\check{Q}_{L(R)}$ are $4\times4$ matrices in the four dimensional
Keldysh$\,\otimes\,$Nambu space. Matrix
$\check{\Xi}=\sigma_{0}\otimes\tau_{z}$, where $\sigma_{i},\tau_{i}$
for $i=0,x,y,z$, are the sets of Pauli matrices acting in the
Keldysh and Nambu subspaces correspondingly, and symbol $\otimes$
stands for the direct product. Matrix $\check{V}$ is the source term
having standard structure in the Keldysh space,
\begin{equation}\label{V}
\check{V}(t)=\left(\begin{array}{cc}V^{cl}(t) & V^{q}(t)
\\ V^{q}(t) & V^{cl}(t)\end{array}\right)\otimes\tau_{0}.
\end{equation}
Diagonal elements of $\check{V}$ are directly related to the
classically applied voltage $V^{cl}(t)=eVt$, while $V^{q}(t)$ is
just its quantum component. This terminology stems from the Keldysh
contour -- terms \textit{classical} and \textit{quantum} imply the
symmetric and anti--symmetric linear combinations of the field
components residing on the forward and backward parts of the Keldysh
contour, respectively.~\cite{Kamenev} Finally, trace operation
$\Tr\{\ldots\}$ in Eq.~\eqref{S-T} assumes summation over the matrix
structure as well as time and spatial integrations. The origin of
phase factors $\exp[\pm i\check{\Xi}\check{V}]$ in Eq.~\eqref{S-T}
is from gauge transformation, which moves different electrochemical
potentials of electrons in the leads from the Green's functions to
the tunneling term. Dynamics of the Green's functions is governed by
the $\sigma$--model action,~\cite{KA,FLS}
\begin{eqnarray}\label{S-sigma}
&&\hskip-.7cm
i\mathbb{S}_{\sigma}[Q_{L},Q_{R}]=
-\sum_{a=L,R}\frac{i\nu_{a}}{2\lambda_{a}}\Tr\left\{\check{\Delta}_{a}
\check{\Sigma}\check{\Delta}_{a}\right\}\\ &-&\sum_{a=L,R}
\frac{\pi\nu_{a}}{4}\Tr\left\{D_{a}\big(\nabla\check{Q}_{a}\big)^{2}-
4\check{\Xi}\partial_{t}\check{Q}_{a}+
4i\check{\Delta}_{a}\check{Q}_{a}\right\}, \nonumber
\end{eqnarray}
where $\nu_{a}$ is the bare normal metal density of states at the
Fermi energy, $D_{a}$ is the diffusion coefficient, $\lambda_{a}$ is
the superconductive coupling constant, and
$\check{\Sigma}=\sigma_{x}\otimes\tau_{0}$. The matrix
superconductive order parameter $\check{\Delta}_{a}(r,t)$ is
\begin{equation}
\check{\Delta}_{a}=\left(\begin{array}{cc}\hat{\Delta}^{cl}_{a}&\hat{\Delta}^{q}_{a}\\
\hat{\Delta}^{q}_{a} & \hat{\Delta}^{cl}_{a}\end{array}\right),\quad
\hat{\Delta}_{a}=\left(\begin{array}{cc}0&\Delta_{a}\\
-\Delta^{*}_{a} & 0\end{array}\right).
\end{equation}
Action \eqref{S-sigma} is subject to the nonlinear constraint
$\check{Q}^{2}_{a}=1$. Physical quantities of interest are obtained
from the action via its functional differentiation with respect to
the appropriate quantum source. For example, tunnel current is found
from the equation
\begin{equation}\label{I-Def}
I(t)=ie\left(\frac{\delta\mathcal{Z}[V]}{\delta
V^{q}(t)}\right)_{V^{q}=0},\quad
\mathcal{Z}[V]=\int\mathbf{D}[Q_{a}]e^{i\mathbb{S}[Q_{L},Q_{R}]},
\end{equation}
where $\mathbb{S}[Q_{L},Q_{R}]=\mathbb{S}_{\sigma}+\mathbb{S}_{T}$.
Corresponding noise power spectrum is defined as
\begin{equation}\label{Noise-Def}
S(\omega)=\int^{+\infty}_{-\infty}
d(t-t')\left(\frac{\delta^{2}\mathcal{Z}[V]}{\delta V^{q}(t)\delta
V^{q}(t')}\right)_{V^{q}=0} e^{-i\omega(t-t')}.
\end{equation}

The procedure of extracting physical observables, outlined above, is
rather general within Keldysh technique. However, for the problem at
hand, information encoded in the actions \eqref{S-T} and
\eqref{S-sigma} is excessive. Indeed, $\mathbb{S}_{\sigma}$
describes not only dynamics of the order parameter $\check{\Delta}$
but also contains explicitly electronic degrees of freedom in the
form of the $\check{Q}$ matrices, which complicates further
analysis. Simplification is possible realizing that dynamics of
$\check{Q}$ is fast as compared to that of $\check{\Delta}$. The
latter is governed by the time scale $\tau_{Q}\sim1/T$, while the
former by $\tau_{\Delta}\sim 1/(T-T_{c})$, and noticeably
$\tau_{\Delta}\gg\tau_{Q}$ when $T\sim T_{c}$. Under this condition,
one may integrate out fast electronic degrees of freedom from action
\eqref{S-sigma} and find an effective theory, which describes space
and time fluctuations of the superconductive order parameter only.
This program was realized for Eq.~\eqref{S-sigma} in the recent
work~\cite{LK} and we will follow here the same route in dealing
with the tunnel term $\mathbb{S}_{T}[V]$.

Let us outline essential elements of the method. Having interest in
the effects of superconductive fluctuations, it is reasonable to
start from the normal metal state with the Green's functions
$\check{Q}_{L(R)}=\check{Q}_{N}$ given by
\begin{equation}
\check{Q}_{N}(\epsilon)=\left(\begin{array}{cc}1^{R}_{\epsilon} & 2F_{\epsilon} \\
0 & -1^{A}_{\epsilon}\end{array}\right)\otimes\tau_{z},\quad
F_{\epsilon}=\tanh\frac{\epsilon}{2T},
\end{equation}
which minimizes action \eqref{S-sigma} for $\check{\Delta}_{a}=0$.
One treats then $\check{\Delta}_{a}$ in perturbation theory on top
of $\check{Q}_{N}$. Technically this program is realized in several
steps. At the first stage one projects $Q$--matrices as
\begin{equation}\label{Q-projection}
\check{Q}_{a}=e^{-i\check{W}^{a}/2}\circ\check{Q}_{N}\circ
e^{i\check{W}^{a}/2}\,,
\end{equation}
where $\check{W}^{a}_{\epsilon\epsilon'}(r)$ carries information
about fast electronic degrees of freedom. Matrix $\check{W}$ is
parametrized by the two complex fields $c_{\epsilon\epsilon'}(q)$
and $\bar{c}_{\epsilon\epsilon'}(q)$ --- Cooper modes, which will be
integrated out eventually. It is convenient to choose
\begin{equation}
\check{W}^{a}=\check{R}\circ\check{\mathcal{W}}^{a}\circ\check{R}^{-1}
\end{equation}
with
\begin{equation}\label{W}
\check{\mathcal{W}}^{a}_{\epsilon\epsilon'}=
(c^{a}_{\epsilon\epsilon'}\tau_{+}+c^{*a}_{\epsilon\epsilon'}\tau_{-})\otimes\sigma_{+}+
(\bar{c}^{a}_{\epsilon\epsilon'}\tau_{+}+\bar{c}^{*a}_{\epsilon\epsilon'}\tau_{-})\otimes\sigma_{-},
\end{equation}
where $\tau_{\pm}=(\tau_{x}\pm i\tau_{y})/2$,
$\sigma_{\pm}=(\sigma_{0}\pm\sigma_{z})/2$ and
\begin{equation}\label{R}
\check{R}_{\epsilon}=\check{R}^{-1}_{\epsilon}=
\left(\begin{array}{cc}1 & F_{\epsilon} \\
0 & -1\end{array}\right)\otimes\tau_{0}.
\end{equation}
One brings then Eq.~\eqref{Q-projection} into action \eqref{S-sigma}
and expands
$\mathbb{S}_{\sigma}[Q_a]\to\mathbb{S}_{\sigma}[W^a,\Delta_a]$ to
the second order in the Cooper modes
$W^{a}=\{c^{a}_{\epsilon\epsilon'},\bar{c}^{a}_{\epsilon\epsilon'}\}$
(details of this procedure are provided in the
Appendix~\ref{sec-Fluctuations}). One finds then that to the leading
order in the coupling $\Tr\{\check{Q}\check{\Delta}\}$, Cooper modes
are connected to the superconductive order parameter according to
the relations
\begin{equation}\label{c-Delta}
c^{a}_{\epsilon\epsilon'}(q)=C^{R}_{\epsilon\epsilon'}(q)
\mathbf{\Delta}^{c}_{\epsilon\epsilon'}(q),\quad
\bar{c}^{a}_{\epsilon\epsilon'}(q)=C^{A}_{\epsilon\epsilon'}(q)
\mathbf{\Delta}^{\bar{c}}_{\epsilon\epsilon'}(q),
\end{equation}
where we have introduced retarded(advanced) Cooperon propagator,
\begin{equation}\label{C}
C^{R(A)}_{\epsilon\epsilon'}(q)=\frac{1}{D_{a}q^{2}\pm
i(\epsilon+\epsilon')},
\end{equation}
and the form factors,
\begin{eqnarray}
&&\mathbf{\Delta}^{c}_{\epsilon\epsilon'}(q)=
-2[\Delta^{cl}_{\epsilon-\epsilon'}(q)+F_{\epsilon}\Delta^{q}_{\epsilon-\epsilon'}(q)],\\
&& \mathbf{\Delta}^{\bar{c}}_{\epsilon\epsilon'}(q)=
\phantom{-}2[\Delta^{cl}_{\epsilon-\epsilon'}(q)-F_{\epsilon'}\Delta^{q}_{\epsilon-\epsilon'}(q)].
\nonumber
\end{eqnarray}
Knowing relations \eqref{c-Delta} Gaussian integration over the
Cooper modes is straightforward,
\begin{equation}
\int\mathbf{D}[W^{a}]\exp(i\mathbb{S}_{\sigma}[W^{a},\Delta_a])
=\exp(i\mathbb{S}_{\mathrm{eff}}[\Delta]).
\end{equation}
The corresponding quadratic form
$\mathbb{S}_{\sigma}[W^{a},\Delta_a]$ should be taken from
Eq.~\eqref{S-c} and one finds as a result,
\begin{equation}\label{S-GL}
\mathbb{S}_{\mathrm{eff}}[\Delta]=\sum_{a=L,R}2\nu_{a}\mathrm{Tr}\left\{\vec{\Delta}^{\dag}_{a}
\hat{\EuScript{L}}^{-1}\vec{\Delta}_{a}\right\},\quad
\vec{\Delta}^{T}_{a}=(\Delta^{cl}_{a},\Delta^{q}_{a}).
\end{equation}
The propagator $\hat{\EuScript{L}}^{-1}(q,\omega)$ governs
superconductive order parameter dynamics. It has typical bosonic
structure in the Keldysh space
\begin{equation}
\hat{\EuScript{L}}^{-1}(q,\omega)=\left(\begin{array}{cc}0 &
\EuScript{L}^{-1}_{A}
\\ \EuScript{L}^{-1}_{R} & \EuScript{L}^{-1}_{K}\end{array}\right),
\end{equation}
with
\begin{eqnarray}\label{L}
&&
\EuScript{L}^{-1}_{R(A)}(q,\omega)=
-\frac{\pi}{8T_{ca}}\big(D_{a}q^{2}+\tau_{\mathrm{GL}}^{-1}\mp
i\omega\big),\\
&& \EuScript{L}^{-1}_{K}(q,\omega)=
B_{\omega}\big[\EuScript{L}^{-1}_{R}(q,\omega)-
\EuScript{L}^{-1}_{A}(q,\omega)\big],\nonumber
\end{eqnarray}
and $\tau_{\mathrm{GL}}=\pi/8(T-T_{ca})$ and
$B_{\omega}=\coth(\omega/2T)$.

Noticeably, effective action \eqref{S-GL} is much simpler than the
original one [Eq.~\eqref{S-sigma}]. However, what is important to
emphasize, is that $\mathbb{S}_{\mathrm{eff}}$ captures correctly
all the relevant low energy excitations of $\Delta_{a}(r,t)$. After
these technical preliminaries we turn now to the applications of the
general formalism based on the effective action
$\mathbb{S}_{\mathrm{eff}}[\Delta]$.

\section{Current noise above $T_c$}\label{sec-Noise}
\subsection{Tunnel current noise}

\begin{figure}
\includegraphics[width=9cm]{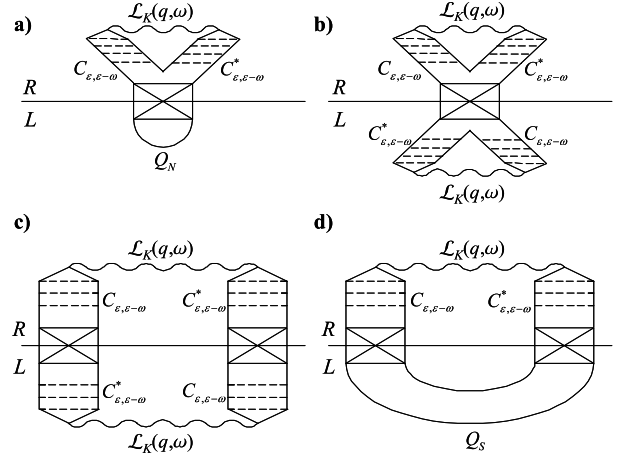}
\caption{Superconductive fluctuation contributions to the current
noise. Diagrams (a) and (b) correspond to the effects coming from
the fluctuations in the density of states for the non-symmetric and
symmetric junctions. Diagrams (c) and (d) are the fluctuating
Josephson current contribution  for the symmetric and
superconductor--fluctuating metal junctions correspondingly. Ladders
represent Cooperons, Eq.~\eqref{C}, wavy lines stand for the
fluctuations propagator, Eq.~\eqref{L}, and crossed boxes depict
tunnel conductance $g_T$. \label{Fig-Diagrams}}
\end{figure}

The first apparent effect of superconductive fluctuations is
modification of the normal metal density of states. Being flat in
the normal state, $\nu(\epsilon)$ acquires strong energy dependence
in the vicinity of $T_c$ with a dip around Fermi
energy.~\cite{Abrahams} The latter suppresses tunnel conductance of
the junction, which influences tunnel current and as the result its
noise. Superconductive fluctuations correction to the tunnel current
was studied in Ref.~\onlinecite{Varlamov-Dorin}. Here we calculate
corresponding correction to the noise. Although the result of this
calculation follows immediately from the fluctuation--dissipation
relation it is still useful to see how it appears within the
$\sigma$--model approach. To this end, assume non--symmetric tunnel
junction: let us say that left electrode is in its normal state,
while the right one is in the fluctuating regime. To calculate noise
power, one uses general definition [Eq.~\eqref{Noise-Def}] and
inserts $\check{Q}_{L}=\check{Q}_{N}$ and
$\check{Q}_{R}\approx\check{Q}_{N}[1+i\check{W}-\check{W}^{2}/2]$
(Ref.~\onlinecite{Note}) into the tunneling part of action
\eqref{S-T}. After the differentiation, which is done with the help
of the formula
\begin{equation}\label{differentiation}
\left.\frac{\delta \exp[\pm i\check{\Xi}\check{V}]}{\delta
V^{q}(t')}\right|_{V^{q}=0}=\pm
i\delta(t-t')\check{\Upsilon}\exp[\pm ieVt\check{\Xi}]\,,
\end{equation}
where $\check{\Upsilon}=\sigma_{x}\otimes\tau_{z}$, one finds for
the noise
\begin{equation}
S(\omega)=S_{S}(\omega)+S_{\mathrm{DOS}}(\omega).
\end{equation}
Here
\begin{equation}\label{Noise-Schottky}
S_{S}(\omega)=2g_{T}T\sum_{\pm}\frac{u_{\pm}}{2T}
\coth\frac{u_{\pm}}{2T}\,,
\end{equation}
with $u_{\pm}=eV\pm\omega$, is just the Schottky formula for the
noise in the normal tunnel junction, while the corresponding
fluctuations correction is
\begin{equation}\label{Noise-Dos-1}
S_{\mathrm{DOS}}(\omega)=\frac{\pi
g_{T}}{8}\int^{+\infty}_{-\infty}d(t-t')
[\mathcal{S}_{+}(t,t')+\mathcal{S}_{-}(t,t')]e^{i\omega(t-t')}\,,
\end{equation}
where
\begin{eqnarray}\label{S-pm}
\mathcal{S}_{\pm}(t,t')=\mathrm{Tr}\left\{\check{Q}_{N}(\epsilon)\check{R}_{\epsilon}
\langle\langle\check{\mathcal{W}}_{\epsilon\epsilon'}(q)
\check{\mathcal{W}}_{\epsilon'\epsilon}(-q)\rangle\rangle\right.\nonumber
\\
\left. \check{R}_{\epsilon}
\check{\Upsilon}\check{Q}_{N}(\epsilon'')\check{\Upsilon} e^{\mp
ieV(t-t')\check{\Xi}}e^{\pm i(\epsilon-\epsilon'')(t-t')}\right\}.
\end{eqnarray}
Quantum averaging in Eq.~\eqref{S-pm}, denoted by the angular
brackets $\langle\langle\ldots\rangle\rangle$, should be performed
with effective action \eqref{S-GL}, namely,
$\langle\langle\ldots\rangle\rangle=\int\mathbf{D}[\Delta]\ldots
\exp\big(i\mathbb{S}_{\mathrm{eff}}[\Delta]\big)$. Recall that
fluctuation matrix $\check{\mathcal{W}}$ is expressed through the
Cooper modes $c_{\epsilon\epsilon'}$ and
$\bar{c}_{\epsilon\epsilon'}$, which are functionally dependent on
the order parameter $\Delta$ via Eq.~\eqref{c-Delta}. The notation
$S_{\mathrm{DOS}}$ in Eq.~\eqref{Noise-Dos-1} and its actual
relation to the density-of-states suppression are motivated in
Appendix~\ref{sec-Noise-Dos}. The linear in $\check{\mathcal{W}}$
term in Eq.~\eqref{S-pm} is not written explicitly since it does not
contribute to the final result. The final comment in order of
Eq.~\eqref{Noise-Dos-1} is that traces of $\mathcal{S}_{\pm}$
functions allow rather simple and convenient diagrammatic
representation shown in Fig.~\ref{Fig-Diagrams}a.

At this point one calculates the product of $\check{\mathcal{W}}$
matrices in Eq.~\eqref{S-pm} and performs Gaussian functional
integration over the fluctuating order parameter using
Eqs.~\eqref{c-Delta} and \eqref{S-GL}. The resulting averages are
\begin{eqnarray}\label{c-averages}
&&\!\!\!\!\!\!\!\!\! \langle\langle c_{\epsilon\epsilon'}(q)
c^{*}_{\epsilon'\epsilon}(-q) \rangle\rangle=(2i/\nu)\nonumber\\
&&\frac{\EuScript{L}_{K}(q,\epsilon-\epsilon')+
F_{\epsilon'}\EuScript{L}_{R}(q,\epsilon-\epsilon')+
F_{\epsilon}\EuScript{L}_{A}(q,\epsilon-\epsilon')}
{\left(Dq^{2}-i(\epsilon+\epsilon')\right)^{2}}\,, \quad \\
 &&\!\!\!\!\!\!\!\!\!\langle\langle\bar{c}_{\epsilon\epsilon'}(q)
\bar{c}^{*}_{\epsilon'\epsilon}(-q) \rangle\rangle=(2i/\nu)\nonumber\\
&&\frac{\EuScript{L}_{K}(q,\epsilon-\epsilon')-
F_{\epsilon'}\EuScript{L}_{A}(q,\epsilon-\epsilon')-
F_{\epsilon}\EuScript{L}_{R}(q,\epsilon-\epsilon')}
{\left(Dq^{2}+i(\epsilon+\epsilon')\right)^{2}}\,.\quad
\end{eqnarray}
Next few steps are conceptually simple. (i) One traces
Eq.~\eqref{S-pm} over its matrix structure first and then performs
time Fourier transforms in Eq.~\eqref{Noise-Dos-1} $\int
d(t-t')e^{i(\epsilon-\epsilon''\pm
eV)(t-t')}=2\pi\delta(\epsilon-\epsilon''\pm eV)$, which removes
$\epsilon''$ integration. (ii) Observe that for the $\epsilon'$
integration, term containing
$F_{\epsilon}\EuScript{L}_{A}(q,\epsilon-\epsilon')$ in the average
$\langle\langle cc^{*}\rangle\rangle$ and term containing
$F_{\epsilon}\EuScript{L}_{R}(q,\epsilon-\epsilon')$ in the average
$\langle\langle \bar{c}\bar{c}^{*}\rangle\rangle$ do not contribute
to $\mathcal{S}_{\pm}$ as being integrals of purely advanced and
retarded functions, respectively. As the result, one takes
$\langle\langle cc^{*}\rangle\rangle+\langle\langle
\bar{c}\bar{c}^{*}\rangle\rangle=2i\mathrm{Im}\langle\langle
cc^{*}\rangle\rangle$. Finally one changes momentum sum into the
integral $\sum_{q}\to\int d^{2}q/4\pi^{2}$, assuming that the
electrodes are quasi--two--dimensional films, and introduces
dimensionless variables $x=Dq^{2}/T$, $y=(\epsilon-\epsilon')/T$,
and $z=(\epsilon+\epsilon')/4T$. After these steps
Eq.~\eqref{Noise-Dos-1} becomes
\begin{eqnarray}\label{Noise-Dos-2}
&&
S_{\mathrm{DOS}}(\omega)=-\frac{16Gi}{\pi^{3}}g_{T}T\sum_{\pm}\coth\frac{u_{\pm}}{2T}\times\\
&&\int^{+\infty}_{0}dx\iint^{+\infty}_{-\infty}dydz\,
\mathrm{Re}\frac{F_{z+u_{\pm}/2T}-F_{z-u_{\pm}/2T}}{\big[(x+\eta)^{2}+y^{2}\big](x+iy-4iz)^{2}}\,.\nonumber
\end{eqnarray}
Here $\eta=1/T_{c}\tau_{\mathrm{GL}}$, and we introduced Ginzburg
number $Gi=1/\nu D$. After the remaining integrations (see
Appendix~\ref{sec-Integrals} for details) one finds as a result,
\begin{eqnarray}\label{Noise-Dos}
S_{\mathrm{DOS}}(\omega)&=&-\frac{4Gi}{\pi^{2}}g_{T}T\ln\left(\frac{T_c}{T-T_c}\right)\\
&\times&\sum_{\pm}
\coth\frac{u_{\pm}}{2T}\mathrm{Im}\psi^{[1]}\left(\frac{1}{2}-\frac{iu_{\pm}}{2\pi
T}\right)\,,\nonumber
\end{eqnarray}
where $\psi^{[1]}(z)$ is the first-order derivative of the digamma
function. Close look on Eq.~\eqref{Noise-Dos} allows us to rewrite
it in the form
\begin{equation}\label{Noise-Dos-FDT}
S_{\mathrm{DOS}}(\omega)=e\sum_{\pm}I_{\mathrm{DOS}}(u_{\pm})\coth\frac{u_{\pm}}{2T}\,,
\end{equation}
where $I_{\mathrm{DOS}}$ is the tunnel current correction calculated
in Ref.~\onlinecite{Varlamov-Dorin}, which is \textit{a priori}
expected result from FDT.

In complete analogy one can calculate corresponding correction to
the noise for the symmetric junction when both electrodes are in the
fluctuating regime. In this case Green's function matrix
$\check{Q}_{L}$ has to be expanded in fluctuations
$\check{\mathcal{W}}$ also and one faces diagram shown in
Fig.~\ref{Fig-Diagrams}b. The result of the calculation can again be
cast in the form of Eq.~\eqref{Noise-Dos-FDT}, where
$I_{\mathrm{DOS}}$ should be replaced by the appropriate
second-order fluctuation correction known from
Ref.~\onlinecite{Varlamov-Dorin}. Furthermore, if one is able to
calculate $I_{\mathrm{DOS}}$ completely, meaning to all orders of
perturbation theory, then for the noise of the tunnel current
Eq.~\eqref{Noise-Dos-FDT} can be considered as the exact result,
which is again consequence of FDT.

\subsection{Josephson current noise}

Contribution to the noise spectrum coming from the Josephson effects
is very much different than that of density of states. First of all
there is no simple FDT relation similar to
Eq.~\eqref{Noise-Dos-FDT}. Secondly, the physical mechanism, which
leads to the noise, is different. Probably the simplest way to see
this is to start from the definition of the current in
Eq.~\eqref{I-Def}. Assuming symmetric junction configuration, one
expands then each Green's--function matrix to the linear order in
fluctuations $\check{Q}_{L(R)}\to i\check{Q}_{N}\check{W}_{L(R)}$ in
the tunnel part of action \eqref{S-T}, which gives for the current,
\begin{equation}\label{I-symm-1}
I_{J}(t)=-\frac{i\pi g_{T}}{4e}\frac{\delta}{\delta
V^{q}(t)}\Tr\left\{e^{i\check{\Xi}\check{V}}\check{Q}_{N}\check{W}_{L}
e^{-i\check{\Xi}\check{V}}\check{Q}_{N}\check{W}_{R}\right\}.
\end{equation}
To proceed further, we will simplify Eq.~\eqref{I-symm-1}, exploring
separation of the time scales between electronic and order parameter
degrees of freedom. Indeed, one should notice that as it follows
from Eq.~\eqref{L} relevant energies and momenta for the
order-parameter variations are
$Dq^{2}\sim\omega\sim\tau^{-1}_{\mathrm{GL}}$, while the relevant
fermionic energies entering the Cooperon in Eq.~\eqref{C} are
$\epsilon\sim\epsilon'\sim1/T$.  As a result, nonlocal relations
between Cooper modes and order parameter in Eqs.~\eqref{W} and
\eqref{c-Delta} can be approximated as~\cite{Remark}
\begin{eqnarray}\label{c-Delta-approx}
&&\hskip-.5cm
\check{\mathcal{W}}^{a}_{tt'}(r)\approx-
\hat{\Theta}_{tt'}\otimes\hat{\Delta}^{a}_{tt'}(r),\quad
\hat{\Theta}_{tt'}= \left(\begin{array}{cc} \theta^{R}_{t-t'} & 0 \\
0 & -\theta^{A}_{t'-t},
\end{array}\right),\nonumber \\
&&\hskip-.5cm
\hat{\Delta}^{a}_{tt'}(r)=\Delta^{cl}_{a}\left(r,\frac{t+t'}{2}\right)\tau_{+}+
\Delta^{*cl}_{a}\left(r,\frac{t+t'}{2}\right)\tau_{-}\,,\qquad
\end{eqnarray}
where $\theta^{R(A)}_{t}$ are the retarded (advanced) step
functions. Physically Eq.~\eqref{c-Delta-approx} implies that
Cooperon is short--ranged, having characteristic length scale
$\xi_{o}=\sqrt{D/T_{c}}$, as compared to the long--ranged
fluctuations of the order parameter, which propagates to the
distances of the order of
$\xi_{\mathrm{GL}}=\sqrt{D\tau_{\mathrm{GL}}}\gg\xi_{o}$. Thus,
relations \eqref{c-Delta} are effectively local, which simplifies
further analysis considerably. Equations \eqref{c-Delta-approx}
allow us to trace Keldysh subspace in Eq.~\eqref{I-symm-1}
explicitly to arrive at
\begin{equation}\label{I-symm-2}
I_{J}(t)=-\frac{\pi
g_{T}}{e}\Tr\left\{\theta_{t_{2}-t_{1}}F_{t_{1}-t}\theta_{t-t_{2}}
\hat{\Delta}^{L}_{tt_{2}}\tau_{z}
\hat{\Delta}^{R}_{t_{2}t_{1}}e^{ieV(t+t_{2})\tau_{z}}\right\}\,,
\end{equation}
where we have used Eq.~\eqref{differentiation} and wrote trace in
the real space representation (note that $\mathrm{Tr}\{\ldots\}$
here does not imply time $t$ integration).  Changing integration
variables $t_{1}=t-\mu$ and $t_{3}=t-\eta$, and rescaling $\eta,\mu$
in the units of temperature $T\eta\to\eta, T\mu\to\mu$, one finds
for Eq.~\eqref{I-symm-2} an equivalent representation,
\begin{eqnarray}\label{I-symm-3}
I_{J}(t)&=&-\frac{i\pi
g_{T}}{eT}\iint\limits^{\quad+\infty}_{-\infty}d\eta d\mu
\frac{\theta_{\eta}\theta_{\mu-\eta}}{\sinh(\pi\mu)}\qquad\qquad\\
&\times& \Tr_{N}\left\{\hat{\Delta}^{L}_{t,t-\frac{\eta}{T}}\tau_{z}
\hat{\Delta}^{R}_{t-\frac{\eta}{T},t-\frac{\mu}{T}}
e^{ieV\big(2t-\frac{\eta}{T}\big)}\right\}\,,\nonumber
\end{eqnarray}
where we used equilibrium fermionic distribution function in the
time domain $F_{t}=-iT/\sinh(\pi Tt)$. The most significant
contribution to the above integrals comes from
$\eta\sim\mu\lesssim1$. At this range ratios $\{\eta,\mu\}/T$ change
on the scale of inverse temperature, while as we already discussed,
order-parameter variations are set by
$t\sim\tau_{\mathrm{GL}}\gg1/T$. Thus, performing $\eta$ and $\mu$
integrations one may neglect $\{\eta,\mu\}/T$ dependence of the
order parameters. As the result we find
\begin{equation}\label{I-symm}
I_J(t)=\frac{i\pi
g_{T}}{4eT}\int\frac{\mathrm{d}^{2}r}{\mathcal{A}}\left[
\Delta^{cl}_{R}(r,t)\Delta^{*cl}_{L}(r,t)e^{-i\omega_{J}t}-c.c\right]\,.
\end{equation}
Finally we are ready to calculate corresponding contribution to the
current noise. One brings two currents from Eq.~\eqref{I-symm} into
Eq.~\eqref{Noise-Def} and pairs fluctuating order parameters using
correlation function,
\begin{equation}\label{Delta-Delta}
\langle\langle\Delta^{cl}_{a}(r,t)\Delta^{*cl}_{b}(r',t')\rangle\rangle_{\Delta}=
\frac{i}{2\nu}\delta_{ab}\EuScript{L}_{K}(r-r',t-t')\,,
\end{equation}
which follows from Eqs.~\eqref{S-GL} and \eqref{L}. As a result,
Josephson current correction to the noise of wide symmetric junction
is
\begin{equation}\label{Noise-symm}
S_{J}(\omega)=-\frac{1}{4\nu^{2}}\left(\frac{\pi
g_{T}}{4eT_{c}}\right)^{2}\sum_{\pm}\int\frac{\mathrm{d}^{2}r}{\mathcal{A}}
\int^{+\infty}_{-\infty}\mathrm{d}t\, \EuScript{L}^{2}_{K}(r,t)\,
e^{-i\omega_{\pm}t},
\end{equation}
where $\omega_{\pm}=\omega\pm\omega_{J}$. Corresponding diagrammatic
representation of Eq.~\eqref{Noise-symm} is shown in
Fig.~\ref{Fig-Diagrams}c. Remaining integrations in
Eq.~\eqref{Noise-symm} can be done in the closed form (see
Appendix~\ref{sec-Integrals} for details), providing
\begin{eqnarray}\label{Noise-symm-fin}
&&\!\!\!\!S_{J}(\omega)=\sum_{\pm}\frac{\pi
Gi^{2}}{64T_{c}}\left(\frac{g_{T}T_{c}}{e}\right)^{2}\frac{\xi^{2}_{o}}{\mathcal{A}}
\left(\frac{T_{c}}{T-T_{c}}\right)^{2}N(\omega_{\pm}\tau_{\mathrm{GL}}),\nonumber\\
&&\!\!\!\!N(z)=\frac{4}{z^{2}}\ln\sqrt{1+z^{2}/4}.
\end{eqnarray}
Analogous calculation in the case of the narrow symmetric junction,
which is obtained from Eq.~\eqref{Noise-symm} by replacing
$\EuScript{L}_{K}(r,t)\to\EuScript{L}_{K}(0,t)$ and removing spatial
integration, gives for the noise spectrum (see details in
Appendix~\ref{sec-Integrals})
\begin{eqnarray}\label{Noise-QPC}
&&\!\!S_{J}(\omega)=\sum_{\pm}\frac{Gi^{2}}{8\pi T_{c}}
\left(\frac{g_{T}T_{c}}{e}\right)^{2}\left(\frac{T_{c}}{T-T_{c}}\right)
M(\omega_{\pm}\tau_{\mathrm{GL}}),\nonumber\\
&&\!\!M(z)=\int^{+\infty}_{1}\frac{\ln(x)\,\mathrm{d}x}{(1+x)^{2}+z^2}.
\end{eqnarray}

\begin{figure}
\includegraphics[width=9cm]{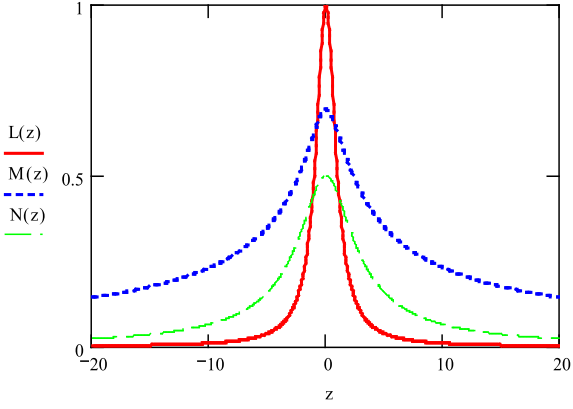}
\caption{(Color online) Shape of the Josephson current noise
spectral lines in the vicinity of the resonances
$z=\omega_{\pm}\tau_{\mathrm{GL}}$. \label{Fig-LMN}}
\end{figure}

In a similar fashion one may consider non--symmetric tunnel
junction. Assume that one of the electrodes is in the deep
superconducting state, with well defined gap in the excitation
spectrum $\Delta_{S}$, while the other is in the fluctuating regime.
We set then one of the $\check{Q}_{a}$ matrices to be
superconductive Green's function $\check{Q}_{L}=\check{Q}_{S}$,
where
\begin{equation}
\check{Q}_{S}=\left(\begin{array}{cc}\hat{Q}^{R}_{S} &
\hat{Q}^{K}_{S} \\ 0 & \hat{Q}^{A}_{S}\end{array}\right),\quad
\hat{Q}^{K}_{S}=\hat{Q}^{R}_{S}\circ
\hat{F}-\hat{F}\circ\hat{Q}^{A}_{S},
\end{equation}
$\hat{F}=F\tau_{z}$, and
\begin{equation}
\hat{Q}^{R(A)}_{S}=\pm\frac{1}{\sqrt{(\epsilon\pm
i0)^{2}-|\Delta_{S}|^{2}}}\left(\begin{array}{cc}\epsilon &
\Delta_{S}
\\ -\Delta^{*}_{S} & - \epsilon\end{array}\right)\, ,
\end{equation}
while expanding the other one in Cooper modes $\check{Q}_{R}\to
i\check{Q}_{N}\check{W}_{R}$. The resulting expression for the
current reads
\begin{equation}
I_J(t)=-\frac{\pi g_{T}}{4e}\frac{\delta}{\delta
V^{q}(t)}\Tr\left\{e^{i\check{\Xi}\check{V}}\check{Q}_{S}e^{-i\check{\Xi}
\check{V}}\check{Q}_{N}\check{W}_{R}\right\}.
\end{equation}
Following the same steps as in the case of the symmetric junction,
carrying out differentiation with the help of
Eq.~\eqref{differentiation} and tracing consequently Keldysh and
Nambu subspaces and performing time integrals, one finds for the
current,
\begin{equation}\label{I-non-symm}
I_J(t)=\frac{i\pi g_{T}}{4e}\ln\left(\frac{|\Delta_{S}|}{T}\right)
\int\frac{\mathrm{d}^{2}r}{\mathcal{A}}
[\Delta^{cl}_{R}(r,t)e^{-i\omega_{J}t}-c.c.],
\end{equation}
where we assumed that $\Delta_{S}\gg T_{c}$. Squaring
Eq.~\eqref{I-non-symm} and averaging over the order-parameter
fluctuations with the help of Eq.~\eqref{Delta-Delta}, we get
\begin{equation}
S_{J}(\omega)=\frac{i}{2\nu}\left(\frac{\pi
g_{T}}{4e}\right)^{2}\ln^{2}\left(\frac{|\Delta_{S}|}{T}\right)
\sum_{\pm}\int\frac{d^{2}r}{\mathcal{A}}\!\!\int^{+\infty}_{-\infty}\!\!\!dt\,
\EuScript{L}_{K}(r,t)e^{-i\omega_{\pm}t}.
\end{equation}
Performing the remaining integrations, one finds noise spectrum of
the non--symmetric junction [see corresponding diagram in
Fig.~\ref{Fig-Diagrams}d],
\begin{eqnarray}\label{Noise-non-symm-fin}
S_{J}(\omega)\!\!&=&\!\!\sum_{\pm}\frac{\pi^{3}Gi}
{64T_c}\left(\frac{g_{T}T_{c}}{e}\right)^{2}
\ln^{2}\left(\frac{|\Delta_{S}|}{T}\right)\frac{\xi^{2}_{o}}{\mathcal{A}}\\
&\times&\!\!\!\left(\frac{T_{c}}{T-T_{c}}\right)^{2}\!\!
L(\omega_{\pm}\tau_{\mathrm{GL}}),\quad
L(z)=\frac{1}{1+z^{2}}.\nonumber
\end{eqnarray}
Spectral line shapes for Eqs.~\eqref{Noise-symm-fin},
\eqref{Noise-QPC} and \eqref{Noise-non-symm-fin} are plotted in the
Fig.~\ref{Fig-LMN}.

\section{Discussions}\label{sec-Summary}

We have considered effects of superconductive fluctuations on the
current noise in tunnel junctions above the critical temperature.
Several contributions were identified. The simplest one originates
from the fluctuation suppression of the density of states. This
effect gives negative contribution to the current noise, which is
only logarithmic in temperature
$S_{\mathrm{DOS}}\propto\ln(T-T_{c})$, whereas dip in the density of
states at the Fermi energy has much stronger temperature dependence
$\delta\nu(0)\propto(T-T_{c})^{-2}$. Somehow current and its noise
get suppressed weaker than the density of states itself. Another
interesting point is that current noise is strongly modified only at
the characteristic voltages $eV\sim T_{c}$, while corresponding
feature in the density of states appears at energies $\epsilon\sim
T-T_{c}$, see Eq.~\eqref{Dos}. It turns out that higher-order
fluctuation effects, similar to that shown in
Fig.~\ref{Fig-Diagrams}b, restore additional structure of the noise
signal at $eV\sim T-T_{c}$. Correction $S_{\mathrm{DOS}}$ is linear
in $Gi$ and in tunnel conductance $g_{T}$. This is in contrast to
the Josephson current contribution to the noise. The latter is
quadratic in fluctuations and in tunneling, and enhances noise at
the frequencies in the vicinity of the Josephson frequency
$\omega_{J}$. The peak at $\omega=\omega_{J}$ is well defined and is
strongly temperature dependent, which makes it possible to detect it
experimentally. We have found that depending on the junction
configuration: symmetric or non-symmetric and narrow or wide, noise
resonance line has different shapes in the frequency domain
Fig.~\ref{Fig-LMN} and different temperature dependencies.

Closing this section we should mention that in the field of
fluctuating superconductivity one usually identifies three types of
fluctuation corrections. Apart from density of states, there are
also so called Aslamazov-Larkin (AL) and Maki-Thompson (MT) terms
mentioned in Sec.~I. It is quite natural to ask how AL and MT
processes modify current noise and how they can be identified within
$\sigma$-model formalism. As an attempt to answer, one should recall
that in addition to the simple tunneling term $\mathbb{S}_{T}[V]$,
considered in this work, one may have yet another one
$i\mathbb{S}_{A}[V]=(\pi
g_{A}/16e^{2})\mathrm{Tr}\{\big(e^{i\check{\Xi}\check{V}}\check{Q}_{L}
e^{-i\check{\Xi}\check{V}}\check{Q}_{R}\big)^{2}\}$, which was
neglected. It corresponds to Andreev processes, and $g_{A}$ is
Andreev conductance. Using $\mathbb{S}_{A}[V]$, instead of
$\mathbb{S}_{T}[V]$, one may follow the same routing expanding
$\check{Q}$-matrices in fluctuations $\check{W}$ to obtain
additional contribution to the noise. However, among all the terms
emerging in perturbative expansion, separation on AL and MT
contributions becomes ambiguous. Nevertheless, the problem is very
interesting and requires further studies.

Many useful discussion with A.~Kamenev, M.~Reznikov, A.~Varlamov and
especially G.~Catelani are kindly acknowledged. I would like to
thank also Digital Material Laboratory at The Institute of Physical
and Chemical Research (RIKEN), where this work was finalized, for
their hospitality. This work was supported by the NSF Grant No.
DMR-0405212 and UMN Graduate School Doctoral Dissertation
Fellowship.

\section{Appendices}\label{sec-Appendix}

\subsection{Fluctuations expansion}\label{sec-Fluctuations}

Within this section we show in details how the transformation from
Eq.~\eqref{S-sigma} to Eq.~\eqref{S-GL} occurs. We start
fluctuations expansion by taking
$\check{Q}\approx\check{Q}_{N}\big(1+i\check{W}-\check{W}^{2}/2\big)$
and bringing it into the $\mathbb{S}_{\sigma}$ (here and in what
follows subscript $a$ in $\check{Q}_{a}$ matrix and all other
elements will be suppressed for brevity). For the trace of the
gradient term we find,
$\Tr\big\{(\nabla\check{Q})^{2}\big\}=-\Tr\big\{\check{Q}_{N}
\nabla\check{W}\check{Q}_{N}\nabla\check{W}\big\}=-
\Tr\big\{\check{\mathcal{W}}\nabla^{2}\check{\mathcal{W}}\big\}$,
where we employed anti--commutativity relation
$[\check{Q}_{N},\check{W}]_{+}=0$ and nonlinear constraint
$\check{Q}^{2}_{N}=1$. Using an explicit form of the
$\check{\mathcal{W}}$ matrix [Eq.~\eqref{W}] and tracing the product
of two $\check{\mathcal{W}}$ over the Keldysh $\otimes$ Nambu space,
we obtain
\begin{equation}\label{Trace-gradient}
\mathrm{Tr}\left\{(\nabla\check{Q})^{2}\right\}=2\mathrm{Tr}\left\{Dq^{2}
\big[c^{*}_{\epsilon\epsilon'}(q)c_{\epsilon'\epsilon}(-q)+
\bar{c}^{*}_{\epsilon\epsilon'}(q)
\bar{c}_{\epsilon'\epsilon}(-q)\big]\right\}\,.
\end{equation}
The time derivative term in the $\mathbb{S}_{\sigma}$ produces
contribution
$\Tr\big\{\check{\Xi}\partial_{t}\check{Q}\big\}=\frac{i}{2}\Tr
\big\{\epsilon\check{R}_{\epsilon}\check{\Xi}
\check{R}_{\epsilon}\check{\Lambda}
\check{\mathcal{W}}_{\epsilon\epsilon'}
\check{\mathcal{W}}_{\epsilon'\epsilon}\big\}$, while linear in
$\check{W}$ part traces out to zero [here we used
$\check{Q}_{N}(\epsilon)=\check{R}_{\epsilon}\check{\Lambda}\check{R}_{\epsilon}$
with $\check{\Lambda}=\sigma_{z}\otimes\tau_{z}$ and substituted
$\partial_{t}\to-i\epsilon$]. Observing that
$\check{R}_{\epsilon}\check{\Xi}
\check{R}_{\epsilon}\check{\Lambda}=\sigma_{z}\otimes\tau_{0}$, one
finds
\begin{equation}\label{Trace-time}
\Tr\left\{\check{\Xi}\partial_{t}\check{Q}\right\}=\frac{i}{2}
\mathrm{Tr}\left\{(\epsilon+\epsilon') [c^{*}_{\epsilon\epsilon'}(q)
c_{\epsilon'\epsilon}(-q)- \bar{c}^{*}_{\epsilon\epsilon'}(q)
\bar{c}_{\epsilon'\epsilon}(-q)]\right\}.
\end{equation}
For the coupling term between Cooperons and $\Delta$, to the leading
order, we have $\Tr\big\{\check{\Delta}\check{Q}\big\}=\Tr\big\{
\check{R}_{\epsilon} \check{\Delta}_{\epsilon-\epsilon'}
\check{R}_{\epsilon'}\check{\Lambda}
\check{\mathcal{W}}_{\epsilon\epsilon'}\big\}$, which translates
into
\begin{equation}\label{Trace-coupling}
\Tr\big\{\check{\Delta}\check{Q}\big\}=-i\mathrm{Tr}\left\{
[\Delta^{cl}_{\epsilon-\epsilon'}(q)+F_{\epsilon}
\Delta^{q}_{\epsilon-\epsilon'}(q)]
c^{*}_{\epsilon'\epsilon}(-q)\right.
\end{equation}
\begin{equation}\nonumber
+[\Delta^{*cl}_{\epsilon-\epsilon'}(q)+
F_{\epsilon}\Delta^{q*}_{\epsilon-\epsilon'}(q)]
c_{\epsilon'\epsilon}(-q)- [\Delta^{cl}_{\epsilon-\epsilon'}(q)-
\end{equation}
\begin{equation}\nonumber
\left.F_{\epsilon'}\Delta^{q}_{\epsilon-\epsilon'}(q)]
\bar{c}^{*}_{\epsilon'\epsilon}(-q)-
[\Delta^{*cl}_{\epsilon-\epsilon'}(q)-F_{\epsilon'}
\Delta^{*q}_{\epsilon-\epsilon'}(q)]
\bar{c}_{\epsilon\epsilon'}(-q)\right\}.
\end{equation}
Combining now Eqs.~\eqref{Trace-gradient}--\eqref{Trace-coupling}
all together and bringing them back into the Eq.~\eqref{S-sigma}, we
wind for the quadratic in Cooperons part of action
$\mathbb{S}_{\sigma}[W^{a},\Delta]=\mathbb{S}^{c}_{\sigma}[c,\Delta]+
\mathbb{S}^{\bar{c}}_{\sigma}[\bar{c},\Delta]$, where contributions
from the retarded $c$ and advanced $\bar{c}$ Cooperons read as
\begin{widetext}
\begin{subequations}\label{S-c}
\begin{equation}
i\mathbb{S}^{c}_{\sigma}[c,\Delta]=-\frac{\pi\nu}{2}\mathrm{Tr}
\left\{c^{*}_{\epsilon\epsilon'}[Dq^{2}-i(\epsilon+\epsilon')]
c_{\epsilon'\epsilon}+2[\Delta^{cl}_{\epsilon-\epsilon'}+
F_{\epsilon}\Delta^{q}_{\epsilon-\epsilon'}]
c^{*}_{\epsilon'\epsilon}+2[\Delta^{*cl}_{\epsilon-\epsilon'}
+F_{\epsilon}\Delta^{*q}_{\epsilon-\epsilon'}]
c_{\epsilon'\epsilon}\right\}\,,
\end{equation}
\begin{equation}
i\mathbb{S}^{\bar{c}}_{\sigma}[\bar{c},\Delta]=-\frac{\pi\nu}{2}\mathrm{Tr}
\left\{\bar{c}^{*}_{\epsilon\epsilon'}[Dq^{2}+i(\epsilon+\epsilon')]
\bar{c}_{\epsilon'\epsilon}-2[\Delta^{cl}_{\epsilon-\epsilon'}-
F_{\epsilon'}\Delta^{q}_{\epsilon-\epsilon'}]
\bar{c}^{*}_{\epsilon'\epsilon}-2[\Delta^{*cl}_{\epsilon-\epsilon'}-
F_{\epsilon'}\Delta^{*q}_{\epsilon-\epsilon'}]
\bar{c}_{\epsilon'\epsilon}\right\}.
\end{equation}
\end{subequations}
\end{widetext}
At this stage we are ready to perform integration over the Cooperon
modes. Assuming that configuration of the order-parameter field is
given, one varies Eq.~\eqref{S-c} with respect to $c^{*}$ and
$\bar{c}^{*}$, and obtains stationary point equations $\delta
\mathbb{S}^{c}_{\sigma}/\delta c^{*}_{\epsilon\epsilon'}=0$ and
$\delta\mathbb{S}^{\bar{c}}_{\sigma}/\delta\bar{c}^{*}_{\epsilon\epsilon'}=0$.
The latter are easily solved by Eq.~\eqref{c-Delta}. Since the value
of the Gaussian integral is equal to that taken at the saddle point,
one brings Eq.~\eqref{c-Delta} into Eq.~\eqref{S-c} and after some
straightforward algebra finds Eq.~\eqref{S-GL}. Further details can
be found in Ref.~\onlinecite{LK}.

\subsection{Relation between $S_{\mathrm{DOS}}$ and $\delta\nu(\epsilon)$}\label{sec-Noise-Dos}

The purpose of this section is to demonstrate explicitly that
$S_{\mathrm{DOS}}$ indeed originates from the DOS effects, which was
hidden in the technical details of Sec.~\ref{sec-Noise}. To this end
we calculate temperature dependence of the $\delta\nu(\epsilon)$
within Keldysh technique. This illustration is useful for the sake
of comparison with the known results obtained previously from the
temperature Matsubara technique~\cite{Abrahams}.

Within $\sigma$--model energy dependent density of states is
expressed in terms of $\check{Q}$ matrix in the following way:
\begin{equation}
\nu(\epsilon)=\frac{\nu}{4}\Tr\big[\check{Q}_{N}
\check{Q}_{\epsilon\epsilon}\big].
\end{equation}
Setting $\check{Q}=\check{Q}_{N}$ one recovers bare normal-metal
density of states $\nu(\epsilon)=\nu$. To account for the
fluctuations on top of the metallic state, one expands $\check{Q}$
in Cooper modes $\check{W}$ to the quadratic order and averages over
$\Delta$ fluctuations with the effective action from
Eq.~\eqref{S-GL};
\begin{equation}
\delta\nu(\epsilon)=-\frac{\nu}{4}\mathrm{Tr} \left[\langle\langle
c_{\epsilon\epsilon'}(q)
c^{*}_{\epsilon'\epsilon}(-q)\rangle\rangle+
\langle\langle\bar{c}_{\epsilon\epsilon'}(q)
\bar{c}^{*}_{\epsilon'\epsilon}(-q)\rangle\rangle\right].
\end{equation}
Observe that this is precisely the same combination of the
Cooperons, which enters $S_{\mathrm{DOS}}$ in the
Eq.~\eqref{Noise-Dos-1}, thus they have common origin. Furthermore,
it is easy to show that $S_{\mathrm{DOS}}(\omega)\propto\int
d\epsilon\,\delta\nu(\epsilon)[F_{\epsilon+u_{\pm}/2T}-F_{\epsilon-u_{\pm}/2T}]$.
Using averages from Eq.~\eqref{c-averages}, density-of-states
correction becomes
\begin{equation}
\frac{\delta\nu(\epsilon)}{\nu}=\mathrm{Im}\sum_{q}
\int^{+\infty}_{-\infty}\frac{d\omega}{2\pi}
\frac{\EuScript{L}_{K}(q,\omega)+F_{\epsilon}\EuScript{L}_{R}(q,\omega)}
{(Dq^{2}-2i\epsilon+i\omega)^{2}}.
\end{equation}
where we set $\epsilon'=\epsilon-\omega$. Here one meets the
convenience of the Keldysh technique, which allows us to get
physical quantities avoiding analytic continuation procedure. Using
explicit form of fluctuations propagators from Eq.~\eqref{L} and
performing frequency and momentum integrations, one finds in the
quasi--two--dimensional case,
\begin{subequations}\label{Dos}
\begin{equation}
\frac{\delta\nu(\epsilon)}{\nu}=-\frac{Gi}{16}
\left(\frac{T_{c}}{T-T_{c}}\right)^{2}\mathcal{F}(\epsilon\tau_{\mathrm{GL}}),
\end{equation}
where dimensionless function is
\begin{equation}
\mathcal{F}(z)=\mathrm{Re}\int^{+\infty}_{0}\frac{d
x}{(1+x)(1+2x-2iz)^{2}}\,.
\end{equation}
\end{subequations}
In agreement with Ref.~[\onlinecite{Abrahams}] dip at the Fermi
energy is $\delta\nu(0)\propto(T-T_{c})^{-2}$, while at large
energies $\epsilon\tau_{\mathrm{GL}}\gg1$ density-of-states
correction recovers its normal value according to
$\delta\nu(\epsilon)\propto(T_{c}/\epsilon)^{2}\ln(\epsilon\tau_{\mathrm{GL}})$.

\subsection{Integrals for $S_{\mathrm{DOS}}(\omega)$ and $S_{J}(\omega)$}\label{sec-Integrals}

(\textbf{I}) Transformation from Eq.~\eqref{Noise-Dos-2} to
Eq.~\eqref{Noise-Dos} requires calculation of the integral,
\begin{equation}
I=\int^{+\infty}_{0}dx\iint^{+\infty}_{-\infty}dydz\,
\mathrm{Re}\frac{F_{z+u_{\pm}/2T}-F_{z-u_{\pm}/2T}}
{\big[(x+\eta)^{2}+y^{2}\big](x+iy-4iz)^{2}}.
\end{equation}
One performs $y$ integration first,
\begin{equation}
I=\pi\int^{+\infty}_{0}dx\int^{+\infty}_{-\infty}dz\,
\mathrm{Re}\frac{F_{z+u_{\pm}/2T}-F_{z-u_{\pm}/2T}}{(x+\eta)(\eta+2x-4iz)^{2}}.
\end{equation}
Since $\eta\ll1$ and relevant $z\sim1$ one may safely approximate
$(\eta+2x-4iz)\approx2(x-2iz)$. Then expanding $F_{z}$ into the
series $F_{z}=2\sum_{n}z/(z^{2}+z^{2}_{n})$, with
$z_{n}=\pi(n+1/2)$, interchanging order of summation and integration
and recalling definition of the $n$th--order derivative of the
digamma function
$\psi^{[n]}(z)=(-1)^{n+1}n!\sum^{\infty}_{n=0}1/(n+z)^{n+1}$, one
finds that
\begin{equation}
\int^{+\infty}_{-\infty}dz\frac{F_{z\pm
u_{\pm}/2T}}{(x-2iz)^{2}}=\frac{i}{2\pi}\psi^{[1]}
\left(\frac{1}{2}\pm\frac{iu_{\pm}}{2\pi T}+\frac{x}{2\pi}\right).
\end{equation}
Remaining $x$ integration can be taken with logarithmic accuracy,
ignoring $x$ dependence of the digamma function since only
$x\lesssim1$ contribute significantly, which eventually gives
$-\ln\eta$. Combining all together, one finds
\begin{equation}
I=\frac{1}{4}\ln(1/\eta)\mathrm{Im}\psi^{[1]}\left(\frac{1}{2}-\frac{iu_{\pm}}{2\pi
T}\right)\,,
\end{equation}
which in combination with Eq.~\eqref{Noise-Dos-2} results in
Eq.~\eqref{Noise-Dos}.

(\textbf{II}) Transition from Eq.~\eqref{Noise-symm} to
Eq.~\eqref{Noise-symm-fin} is performed in the following way. As the
first step one finds Keldysh component of the fluctuation propagator
in the mixed momentum/time representation
$\EuScript{L}_{K}(q,t)=\int\EuScript{L}_{K}(q,\omega)e^{-i\omega
t}d\omega/2\pi$, which gives
\begin{equation}
\EuScript{L}_{K}(q,t)=-\frac{2iT^{2}_{c}}{T-T_{c}}
\frac{e^{-\varkappa_{q}|t|/\tau_{\mathrm{GL}}}}{\varkappa_{q}},\quad
\varkappa_{q}=(\xi_{\mathrm{GL}}q)^{2}+1.
\end{equation}
One inserts then
$\EuScript{L}_{K}(r,t)=\int\EuScript{L}_{K}(q,t)e^{iqr}dq^{2}/4\pi$
into Eq.~\eqref{Noise-symm}, integrates over $r$, introduces
dimensionless time $\tau=t/\tau_{\mathrm{GL}}$, and changes from $q$
to $\varkappa$ integration
$dq^{2}=d\varkappa/\xi^{2}_{\mathrm{GL}}$, which gives all together
\begin{eqnarray}
S_{J}(\omega)=\sum_{\pm}\frac{\pi
Gi^{2}}{64T_c}\left(\frac{g_{T}T_c}{e}\right)^{2}\frac{\xi^{2}_{o}}{\mathcal{A}}
\left(\frac{T_{c}}{T-T_{c}}\right)^{2}\qquad\qquad\nonumber\\
\times\int^{+\infty}_{-\infty}d\tau\int^{+\infty}_{1}
\frac{d\varkappa}{\varkappa^{2}}e^{-2\varkappa|\tau|-iz_{\pm}\tau}\,,
\end{eqnarray}
where $z_{\pm}=\omega_{\pm}\tau_{\mathrm{GL}}$. After $\tau$
integration one is left with
\begin{equation}
\int^{\infty}_{1}\frac{4d\varkappa}{\varkappa(4\varkappa^{2}+z^{2}_{\pm})}\,,
\end{equation}
which defines $N(z)$ function in Eq.~\eqref{Noise-symm-fin}.

(\textbf{III}) Calculation of Eq.~\eqref{Noise-QPC} is completely
analogous. Noticing that
$\EuScript{L}_{K}(0,t)=\int\EuScript{L}_{K}(q,t)dq^{2}/4\pi$ and
transforming to the dimensionless units $\tau=t/\tau_{\mathrm{GL}}$
and $\varkappa=(\xi_{\mathrm{GL}}q)^2+1$, we have
\begin{eqnarray}
S_{J}(\omega)=\sum_{\pm}\frac{Gi^{2}\tau_{GL}}{4\pi^{2}}\left(\frac{g_{T}T_c}{e}\right)^{2}
\qquad\qquad\qquad\nonumber\\\times
\int^{+\infty}_{-\infty}d\tau\iint^{+\infty}_{1}\frac{d\varkappa
d\varkappa'}{\varkappa\varkappa'}e^{-(\varkappa+\varkappa')|\tau|-iz_{\pm}\tau}.
\end{eqnarray}
After $\tau$ integration one is left with
\begin{equation}
\iint^{+\infty}_{1}\frac{d\varkappa
d\varkappa'}{\varkappa\varkappa'}\frac{\varkappa+\varkappa'}{(\varkappa+\varkappa')^{2}+z^{2}_{\pm}}
=\frac{2}{z_{\pm}}\int^{+\infty}_{1}\frac{d\varkappa}{\varkappa}\mathrm{arccot}
\left(\frac{1+\varkappa}{z_{\pm}}\right)
\end{equation}
which after the integration by parts reduces to $2M(z)$, with $M(z)$
function defined by Eq.~\eqref{Noise-QPC}.



\begin{thebibliography}{99}
\bibitem{Note} Here additional subscript $R$ in the notations of matrix
$\check{W}^{R}$ was suppressed for brevity. According to the
assumption $\check{W}^{L}=0$ for the non-symmetric configuration.
\bibitem{Remark} Deriving Eq.~\eqref{c-Delta-approx} we have ignored quantum
component of the order parameter $\Delta^{q}(r,t)$. Although formaly
present it gives at the end subleading contribution to the noise
spectrum.
\end{thebibliography}
\end{document}